# Network analysis reveals distinct clinical syndromes underlying acute mountain sickness.

David P Hall[1,2], Ian JC MacCormick[2], Alex T Phythian-Adams[2], Nina M Rzechorzek[2,3], David Hope-Jones[2], Sorrel Cosens[2], Stewart Jackson[2], Matthew GD Bates[2,4], David J Collier[5], David A Hume[6], Thomas Freeman[6], AA Roger Thompson[2,7], J Kenneth Baillie[2,6]

1. RAF Centre of Aviation Medicine, RAF Henlow, Beds, SG16 6DN, United Kingdom

2. Apex (Altitude Physiology Expeditions), c/o Dr JK Baillie, Critical Care Medicine, University of Edinburgh, Royal Infirmary of Edinburgh, 54 Little France Drive, EH16 4SA, United Kingdom

3. Centre for Clinical Brain Sciences, University of Edinburgh, Edinburgh, United Kingdom

4. Institute for Ageing and Health, Newcastle University, Newcastle upon Tyne, United Kingdom

5. William Harvey Research Institute, Queen Mary University of London, London, UK

6. Division of Genetics and Genomics, Roslin Institute, Edinburgh, United Kingdom

7. Academic Unit of Respiratory Medicine, Department of Infection and Immunity, University of Sheffield, Sheffield, United Kingdom

Address for correspondence: Dr J Kenneth Baillie, Clinical Lecturer, Critical Care Medicine, University of Edinburgh, Royal Infirmary of Edinburgh, 54 Little France Drive, EH16 4SA, United Kingdom

Tel: +44 (0) 131 651 9204

email: j.k.baillie@ed.ac.uk

Running title: AMS: a diverse spectrum of clinical syndromes

Manuscript Word Count: 2772

Abstract Word Count: 250

This article has an online supplementary data which is available at http://www.altitude.org/

## Contributions of authors

JKB conceived and designed the study and led the computational analysis; DPH, TF and JKB performed computational analyses; IJCM, ATP-A, NR, DH-J, SC, SJ, MGDB, AART and JKB gathered the data; DJC designed the questionnaire; DPH, AART and JKB wrote the manuscript. All authors contributed to data interpretation and approved the final manuscript.

## Abstract

Acute mountain sickness (AMS) is a common problem among visitors at high altitude, and may progress to life-threatening pulmonary and cerebral oedema in a minority of cases. International consensus defines AMS as a constellation of subjective, non-specific symptoms. Specifically, headache, sleep disturbance, fatigue and dizziness are given equal diagnostic weighting. Different pathophysiological mechanisms are now thought to underlie headache and sleep disturbance during acute exposure to high altitude. Hence, these symptoms may not belong together as a single syndrome. Using a novel visual analogue scale (VAS), we sought to undertake a systematic exploration of the symptomatology of AMS using an unbiased, data-driven approach originally designed for analysis of gene expression. Symptom scores were collected from 293 subjects during 1110 subject-days at altitudes between 3650m and 5200m on Apex expeditions to Bolivia and Kilimanjaro. Three distinct patterns of symptoms were consistently identified. Although fatigue is a ubiquitous finding, sleep disturbance and headache are each commonly reported without the other. The commonest pattern of symptoms was sleep disturbance and fatigue, with little or no headache. In subjects reporting severe headache, 40% did not report sleep disturbance. Sleep disturbance correlates poorly with other symptoms of AMS (Pearson r = 0.31 vs headache). These results challenge the accepted paradigm that AMS is a single disease process and describe at least two distinct syndromes following acute ascent to high altitude. This approach to analysing symptom patterns has potential utility in other clinical syndromes.

## Introduction

Acute mountain sickness (AMS) occurs in up to 50% of individuals ascending to high altitude (Maggiorini et al., 1990) and may progress to life-threatening pulmonary and cerebral oedema in a minority of cases (Roach, and Hackett, 2001). The present international consensus defines AMS as a collection of subjective, non-specific symptoms(Roach et al., 1993). Specifically, headache, sleep disturbance, and vague symptoms of fatigue and dizziness and given equal diagnostic weighting. Since we lack a common underlying mechanism to explain these symptoms, it is far from certain that they belong together as a single syndrome.

The most frequently-used criteria for the definition of AMS are based on the self-reported Lake Louise consensus scoring system (LLS) (Roach et al., 1993). A positive Lake Louise Score can describe a spectrum of non-specific symptoms experienced on exposure to high altitude, and these may encompass more than one disease phenotype (West, 2011).

Headache is the cardinal feature of the LLS and is required under the present criteria for diagnosis of AMS (Roach et al., 1993). There is evidence to suggest that the development of mild vasogenic cerebral oedema leading to increased intracranial pressure may be an important factor in the development of high altitude headache and AMS (Kallenberg et al., 2007). Optic nerve sheath diameter, an indirect measure of intracranial pressure, increases with altitude and also correlates with AMS score (Fagenholz et al., 2009; Sutherland et al., 2008). Several small case series in which participants were subjected to simulated high altitude before undergoing cerebral MRI also demonstrated an increase in brain volume (Kallenberg et al., 2007; Morocz et al., 2001). A related hypothesis proposes that fluid redistribution to the intracellular space leading to astrocytic swelling underlies the development of symptomatic AMS (Bailey et al., 2009).

In contrast, sleep disturbance at high altitude may be a mechanistically separate problem. Changes in the control of ventilation following acute ascent to altitude lead to a cyclical respiratory pattern. The hypoxic ventilatory drive causes hypocapnia and a reduction in respiratory drive (Wilson et al., 2009). During sleep, in the absence of the wakefulness drive to breathing, hypoventilation results in cyclical hypoxia and recurrent interruption of deep sleep stages (Burgess et al., 2004).

As a categorical score, LLS is inherently disadvantaged in any attempt to quantify a continuous spectrum of disease severity. A further problem in the measurement of AMS is that LLS scores at high altitude produce a strongly skewed dataset, reducing the power of statistical analyses and making it difficult to quantitatively compare different symptoms, or to infer relationships between severity and other physiological or biochemical measurements.

Visual analogue scales (VAS) have been rigorously validated as tools to quantify the severity of subjective symptoms, particularly in pain medicine (DeLoach et al., 1998; Gallagher et al., 2001; Price et al., 1983; Todd et al., 1996), and are increasingly accepted for symptom scoring at altitude (Roach, and Kayser, 2007). Harris *et al.* used VAS scores as an endpoint in a randomised controlled trial comparing treatments of high altitude headache (Harris et al., 2003). Since then, there have been four studies exploring the relationship between VAS and LLS describing altitude symptomatology (Hext et al., 2011; Kayser et al., 2010; Van Roo et al., 2011; Wagner et al., 2007), consistently demonstrating a correlation between LLS and VAS, both for individual components of LLS and overall LLS score. However, both the linearity of the relationship and the threshold VAS measurement for diagnosis of AMS are poorly defined. Unlike LLS, which is a standardised score and comparable across different studies, VAS studies have used different scores and as yet there is no consensus VAS for altitude illness.

We sought to perform an unbiased quantitative assessment of the symptomatology of AMS in healthy lowlanders acutely exposed to high altitude. We used a novel VAS questionnaire to record subjective symptoms experienced by research subjects over the course of 1110 subject-days at high altitude. Here we report the comparison between VAS and LLS, and apply a hypothesis-free clustering methodology widely used in transcriptomics (Hume et al., 2010) to identify different patterns of symptoms in individuals acutely exposed to high altitude.

## Materials and Methods

### Study Population

Data were collected from subjects at high altitude during two research expeditions. 103 participants were recruited from the Apex 2 expedition to the Bolivian Andes (Baillie et al., 2009; Bates et al., 2011; Thompson et al., 2006). Subjects flew to La Paz, Bolivia (3650 m) and after 4–5 days' acclimatisation ascended over a period of 90 minutes to 5200 m by off-road vehicle. 41 subjects were randomised to receive antioxidant supplements (1 g L-ascorbic acid, 400 IU alpha-tocopherol acetate, and 600 mg alpha-lipoic acid/day in four divided doses starting five days before ascent to 5200 m), 20 subjects were randomised to receive sildenafil (150 mg daily in three divided doses) and the remaining 42 received placebo only. This study was approved by the Lothian Research Ethics Committee. Data were also included from 189 trekkers ascending to the Kibo Hut at 4730 m on Mt Kilimanjaro in Tanzania. This study received ethical approval from the Tanzanian Institute for Medical Research and is described elsewhere (Jackson et al., 2010). Of 189 subjects, 149 were attempting to summit via the standard Marangu route, which involves sleeping at huts located at 2743 m, 3760 m and 4730 m on the 5895 m summit attempt. Of these, 82 climbers had taken an additional rest day at 3760 m. The remaining 40 subjects were completing the Rongai route with no fixed sleeping points.

**Measurement of symptoms**

Subjects were asked to complete a novel seven-question visual analogue scale (VAS) questionnaire during acute exposure to high altitude (conceived and designed by DJC, and provided in the Supplementary figure S1). This assesses five symptoms associated with high altitude: headache ("no headache at all" to "worst headache ever"), nausea (1. "I really want to be sick" to "don't feel sick at all"; 2."my guts are really bad" to "my guts are fine"), fatigue (1. "I'm totally exhausted" to "I'm full of energy"; 2. "I'm at my best" to "I'm at my worst"), dizziness ("no dizziness at all" to "more dizzy than ever") and sleep quality ("worst night's sleep ever" to "best night's sleep ever"). Subjects were asked to score their symptoms on a spectrum between the two extreme statements by marking their position on a horizontal 100 mm line. On all questionnaires, the direction of the scales was reversed for several questions to ensure subjects paid close attention to the questions. Reversed VAS scales were corrected prior to analysis, such that 0mm corresponded to minimal severity and 100 mm to maximal severity for all seven questions. Summing the seven components produced a maximum total VAS score of 700 mm. The VAS included a quality control question, which asked with different wording about symptoms of fatigue in order to detect inattentive subjects. A minimum agreement of 40 mm was required between the responses to two questions ("I'm totally exhausted" to "I'm full of energy", and "I'm at my best" to "I'm at my worst").

Subjects also completed the Lake Louise Score (LLS) questionnaire, which includes assessment of headache, nausea and vomiting, fatigue/weakness, dizziness and sleep quality. Participants score the severity of each symptom from 0 (no symptom) to three (maximal severity); a total score of greater than or equal to three with the presence of headache equates to acute mountain sickness (Roach et al., 1993).

Sequential LLS and VAS data were collected daily from each of the 103 Apex 2 subjects over 13 days of exposure to high altitude (5 days at 3600 m, 8 days at 5200 m). Cross-sectional data from the 189 Kilimanjaro subjects were recorded at 4730 m during their ascent.

**Data Analysis**

VAS scores were measured by hand in mm. Groups of VAS questionnaires exhibiting similar symptom profiles were identified using a network analysis tool, BioLayout Express 3D, http://www.biolayout.org/ (Theocharidis et al., 2009). This software is primarily used to visualize and cluster networks from microarray expression data (Freeman et al., 2007). We created an undirected network, in which each node (depicted by coloured spheres) represented one VAS questionnaire. Weighted edges (interconnecting lines) between these nodes represent Pearson correlation coefficients (r) between the symptoms in each questionnaire above a threshold of r=0.95. The MCL clustering algorithm (inflation = 1.4) was used to subdivide the network into discrete clusters of VAS questionnaires sharing similar features (van Dongen, 2000).

The relationship between symptoms was explored by generating a correlation matrix in Microsoft Excel between VAS scores for each of five symptoms, and in Biolayout Express 3D. The correlation between VAS and LLS was explored using Spearman's rank co-efficient, calculated in R 2.13.1 (R Foundation for Statistical Computing, Vienna, 2011).

A simple graphical interface was developed (Excel 2003, Microsoft Corp, Redmond, WA), to allow exploration of the effects of changing the weighting given to each symptom recorded on the VAS score, the degree of concordance required between the to quality control questions and the effects of square-root normalization on the LLS and VAS data. The full dataset and formulae for data manipulation are provided as supplementary information at www.altitude.org/ams.php

## Results

Visual analogue scale (VAS) and Lake Louise Score (LLS) data were collected from 292 individuals over the course of 1110 subject-days at high altitude. This included 189 questionnaires from 189 subjects at the Kibo Hut (4730 m) on Mt Kilimanjaro and 921 serial questionnaires from 103 subjects on the Apex 2 expedition. 65 questionnaires (13 from Kilimanjaro, 52 from Apex 2, 5.9% of total questionnaires) were excluded as they did not meet the minimum agreement threshold of 40 mm between the two quality control questions on the VAS questionnaire. This left a total of 1045 subject-days at altitude (176 at Kilimanjaro, 869 on Apex 2), with both corresponding VAS and LLS data.

**Cluster Analysis of AMS Symptoms**

In order to maximise the power to detect different patterns of associated symptoms, we chose a composite analysis, including all questionnaires from within 1 week of a recent increase in altitude. This combined data from disparate sources, with differences in mode and rate of ascent, geographical location and drug treatment, and includes several questionnaires from the same volunteers at different times. The patterns reported here are consistently replicated in subsets of volunteers with same location and ascent rate (Supplementary figure S4), from the placebo group only and including only one questionnaire for each volunteer (Supplementary figure S3).

In order to explore different patterns of related symptoms, VAS data for the 1045 subjects-days at altitude were analysed in BioLayout Express 3D. The network generated was clustered using the MCL clustering algorithm (inflation value 1.4, minimum cluster size 30 nodes) producing three main clusters (Figure 3). The largest cluster contained 407 questionnaires, and corresponded to subjects who slept poorly, were fatigued, but had minimal headache or other symptoms. The second cluster of 127 questionnaires, contained subjects who reported, poor sleep, headache, and fatigue. The third cluster contained 43 questionnaires and corresponded to subjects who had little sleep disturbance, but had headache and fatigue. The remaining questionnaires did not form a cluster large enough to be included in the analysis – the full dataset may be explored at www.altitude.org/ams.php. No relationship was seen between age or sex and any pattern of symptoms (Figure 4).

**Symptom Correlation**

The relationship between different symptoms (rather than symptom profiles) was also investigated using the correlation between two symptoms across the whole population of responses. Symptoms as recorded by the visual analogue scale inter-correlated as shown in Table 1. Sleep was most weakly correlated with the five other symptoms (mean correlation co-efficient 0.33), whereas fatigue was most strongly correlated with the other reported symptoms (mean correlation co-efficient 0.55). A network showing the relationship between symptoms is shown in Figure 5.

**Timing of sleep assessment**

For practical reasons, AMS questionnaires inquire about sleep quality on the night *preceding* the questionnaire. This is one reason why AMS scores are lower on the first day following acute ascent. However in physiological terms it is equally logical to assess sleep on the *following* night. This is particularly important when assessing the correlation of sleep with other symptoms, because the better sleep quality experienced on the night before ascent will cause an artefactual dissociation between sleep quality and the other symptoms of AMS.

Our dataset enables us to address the effect of this change. We repeated the correlation analyses on a subset of questionnaires in which we substituted the sleep score from the following day. For both the clustering of questionnaires with similar patterns of symptoms (Supplementary figure S3), and the correlation between the 5 symptoms across all volunteers, the patterns described above were consistently replicated.

**Quantification of severity**

384 of the 1045 questionnaires were consistent with a diagnosis of acute mountain sickness using the Lake Louise Score (LLS > 2 with LLS headache > 0). One subject developed high altitude pulmonary edema (HAPE) at 3600m; there were no cases of HACE (high altitude cerebral edema). The median of all LLS scores was 2 (interquartile range 1-4). The mean total VAS was 192.6 mm with a standard deviation of 124.3 mm. A square-root transformation resulted in normalisation of the VAS data (Kolmogorov-Smirnov test: KS distance 0.02036, P>0.10) but not the LLS data (Kolmogorov-Smirnov test: KS distance 0.1620, P<0.0001) (Figure 2).

## Discussion and Conclusion

The scale of the dataset collated here, collected during 1110 subject-days at altitude, enables for the first time a comprehensive quantitative analysis of symptom correlation in altitude illness. We show that sleep disturbance is an outlier, correlating poorly with other symptoms of AMS. By identifying clusters sharing common patterns of symptoms, we demonstrate that distinct patterns of disease are primarily characterised by the presence of sleep disturbance or headache.

Among symptomatic individuals following acute ascent to high altitude, some degree of fatigue was a ubiquitous finding. In contrast, sleep disturbance and headache are each commonly reported without the other. This is consistent with the hypothesis that distinct pathogenetic mechanisms underlie sleep disturbance and headache at high altitude.

Analysis of correlation networks is a standard tool for drawing biological meaning from transcriptomic data (Hume et al., 2010; Mabbott et al., 2010; Summers et al., 2010). As we demonstrate here, the same approach may also have substantial utility in classifying disease phenotypes. By providing an unbiased allocation of subjects with similar symptoms into clusters, it provides a novel and informative method for identifying patterns of disease. This method may easily be extended to other syndromes.

Our primary dataset contains repeated measures (1110 questionnaires obtained from 292 subjects), which may lead to a learning bias in subjects completing the VAS (Van Roo et al., 2011). We also used symptom scores from some subjects who were taking oral antioxidant supplements or sildenafil for other trials. However these limitations do not affect the main outcomes. Repeat analyses focussing only on a single time point, or removing subjects on active drug treatments, did not affect the results.

Our study makes use of the largest VAS dataset to date to explore altitude-associated symptomatology. As expected, there is a strong correlation between VAS and LLS (Hext et al., 2011; Kayser et al., 2010; Van Roo et al., 2011; Wagner et al., 2007). Our work extends previous attempts to evaluate VAS scores by demonstrating the utility of VAS in creating a normally-distributed, continuous measure of AMS. VAS have a high test-retest reliability and inter-rater reliability (Roach, and Kayser, 2007) and may be more sensitive than discrete measures (Grant et al., 1999). In future research using VAS as an endpoint or covariate, improved statistical power may be expected from the use of parametric tests that may only be applied to normally-distributed, continuous variables.

Furthermore, combining distinct pathophysiological processes into a single summary measure carries a risk of introducing excessive noise into studies of the pathogenesis of AMS, and of reducing the magnitude of effect signals in therapeutic studies. Focussing only on a group of well-correlated symptoms that share common mechanisms will further improve the quality of future research into AMS.

To be useful for clinical and physiological studies, a symptom score must accurately reflect the severity of a single clinical entity (Bartsch et al., 2004). It is therefore critical that all of the symptoms included share an underlying pathogenesis. If not, effective treatments will be falsely rejected, and attempts to understand disease processes will be obstructed. It is our view that the results presented here should provoke a reassessment of consensus diagnostic criteria and severity measures for AMS.

## Financial Disclosure

All authors report no support from any organisation for the submitted work. No external funders had any role in study design, data collection and analysis, decision to publish, or preparation of the manuscript. All authors declare that no competing interests exist.

## Acknowledgments

We would like to thank the participants of the Apex 2 expedition and the trekkers who volunteered as subjects whilst ascending Mt Kilimanjaro.

# Figures

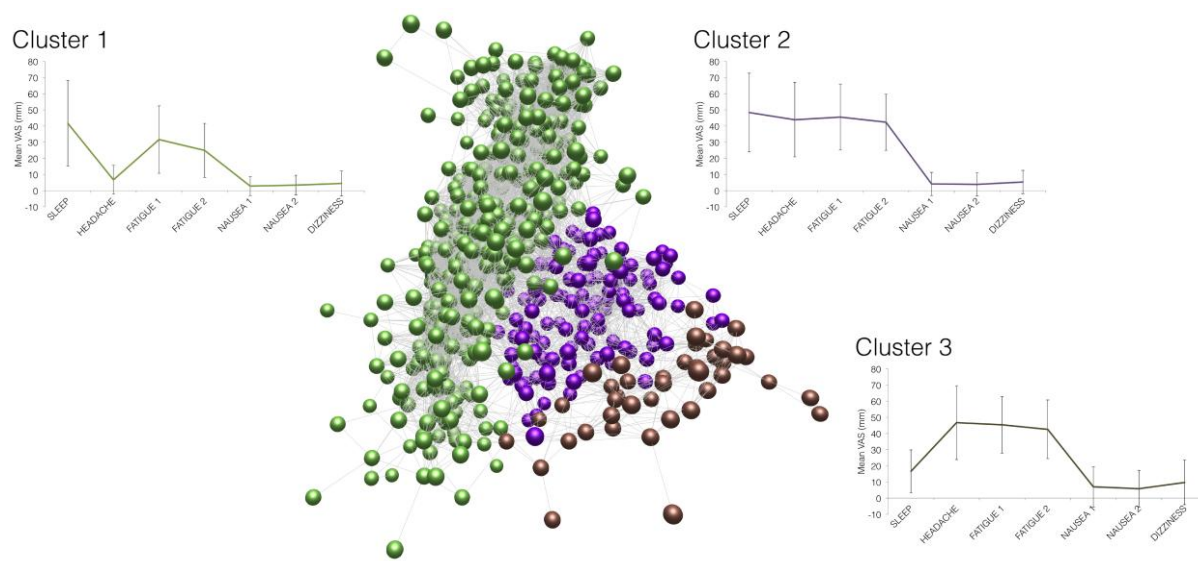

**Figure 1. Identification of VAS questionnaires exhibiting similar symptom profiles using Biolayout Express 3D.**

Each node (coloured sphere) represents a VAS questionnaire. Nodes are connected by weighted lines, which represent correlations between similar symptom profiles. Nodes are connected with each other if the Pearson correlation coefficient between them exceeds 0.95. The MCL clustering algorithm (inflation = 1.4) sub-divided this network into three discrete clusters of VAS questionnaires, each of which shared similar features. Figures adjacent to the clusters represent the median VAS scores for each question in the VAS questionnaire. The green cluster (cluster 1) contains 407 nodes and corresponds to subjects who slept poorly, and were fatigued but had little headache. The brown cluster (cluster 2) contains 127 nodes and corresponds to subjects who slept poorly and did have headache. The purple cluster (cluster 3) contains 43 nodes and corresponds to subjects who had little sleep disturbance but had headache. The remaining nodes do not correlate sufficiently with each other to form a significant cluster.

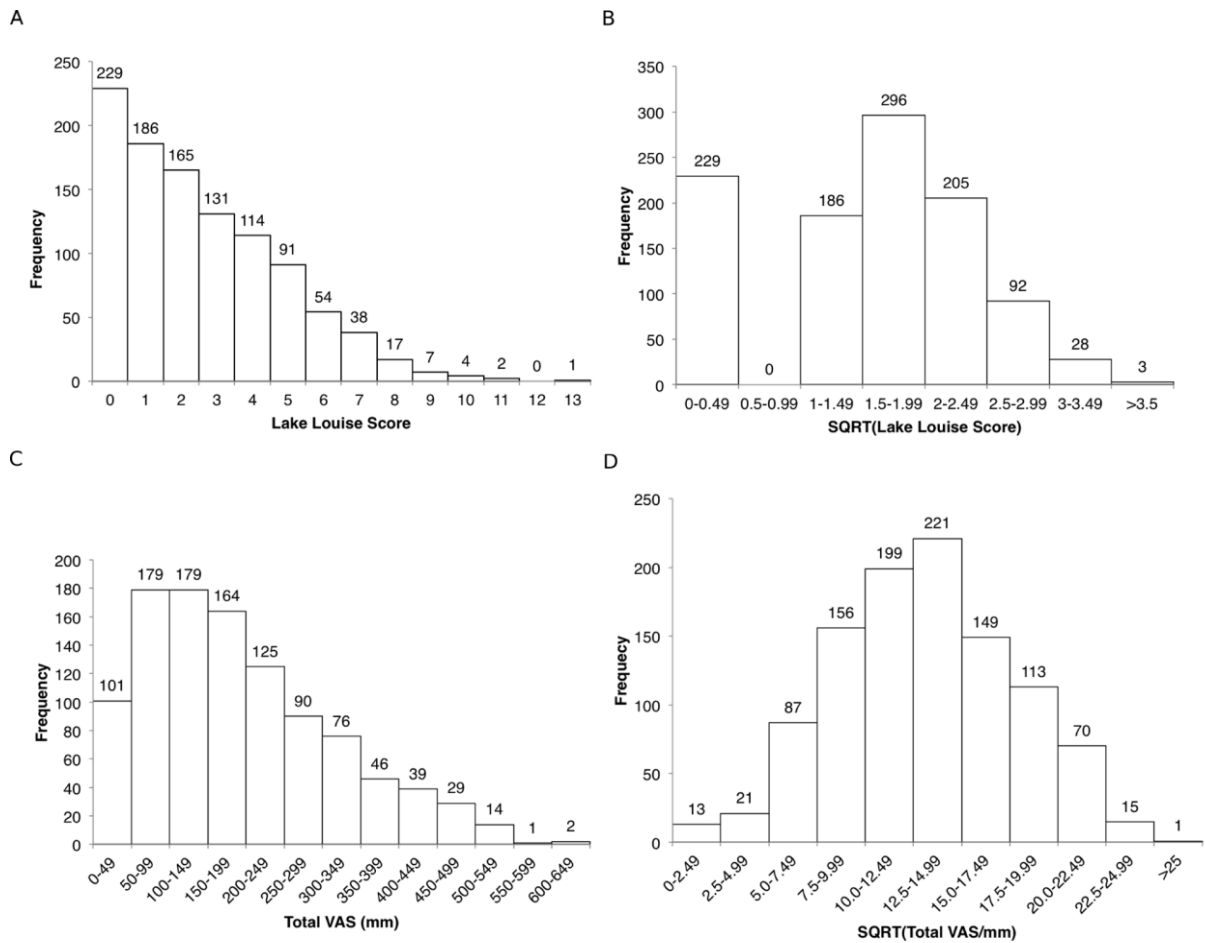

**Figure 2. Frequency distribution of LLS and VAS scores (n=1045).**

(A) Distribution of LLS. A positive LLS, indicating AMS, is a score of 3 or greater in the presence of headache; (B) Distribution of Lake Louise Scores following square-root transformation; (C) Distribution of total VAS scores (minimum 0mm; maximum 700mm); (D) Distribution of total VAS scores following square-root transformation of data. LLS: Lake Louise Score; VAS: visual analogue scale; AMS: acute mountain sickness.

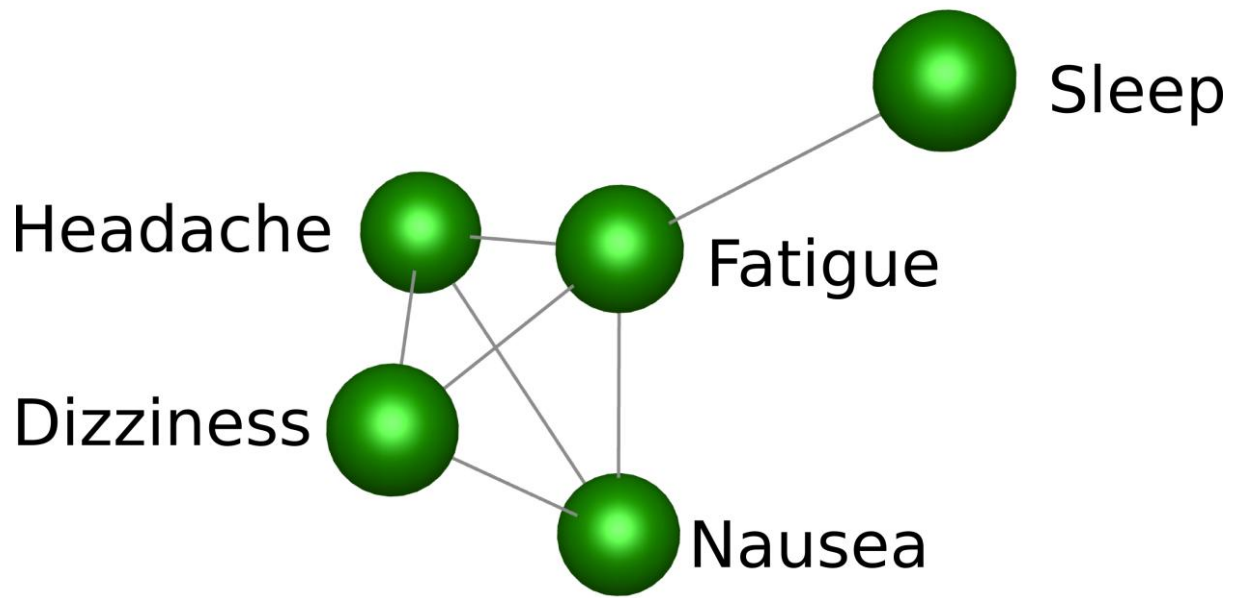

**Figure 3. Correlations between different LLS symptoms.**

The correlations between symptoms included in the Lake Louise Score was explored across the whole population of responses (n=1045) using Biolayout 3D (minimum Pearson correlation cut–off r=0.4). Headache, fatigue, nausea and dizziness all correlate with each other, whereas sleep is an outlier and correlates only with fatigue in this model.

**Table 1**

|          | Sleep | GI upset | Dizziness | Headache | Fatigue |
|----------|-------|----------|-----------|----------|---------|
| Sleep    | x     | 0.28     | 0.26      | 0.31     | 0.45    |
| GI upset | 0.28  | x        | 0.48      | 0.42     | 0.53    |
| Dizziness| 0.26  | 0.48     | x         | 0.52     | 0.51    |
| Headache | 0.31  | 0.42     | 0.52      | x        | 0.61    |
| Fatigue  | 0.45  | 0.53     | 0.51      | 0.61     | x       |
| *Average*| *0.33*| *0.48*   | *0.50*    | *0.51*   | *0.55*  |

**Pearson correlation coefficients between VAS scores for the different symptom components of the LLS.** Repeat measures (for GI upset and fatigue) were averaged.

## Supplementary Figures

**Visual Analog Scales**

Please use a pen to mark a single vertical line anywhere on the horizontal lines below at the point which best describes how you feel.

NAME:………………………………………..

SUBJECT NO.: ……………………………….

DATE: ………………………………………..

LA PAZ DAY (CIRCLE):     1     2     3     4

CHACALTAYA DAY (CIRCLE):
1     2     3     4     5     6     7     8     9

Worst night's sleep ever ─────────── Best night's sleep ever

No headache at all ─────────── Worst headache ever

I really want to be sick ─────────── Don't feel sick at all

I'm at my best ─────────── I'm at my worst

My guts are really bad ─────────── My guts are fine

I'm totally exhausted ─────────── I'm full of energy

No dizzyness at all ─────────── More dizzy than ever

*David Collier, William Harvey Research Institute, St Bart's & The London, Queen Mary School of Medicine & Dentistry, Londo*

**S1. VAS questionnaire form.**

Questionnaire given to subjects to record VAS scores relating to symptoms experienced at altitude on the Apex 2 expedition. The same form was used by participants on the Kilimanjaro expedition.

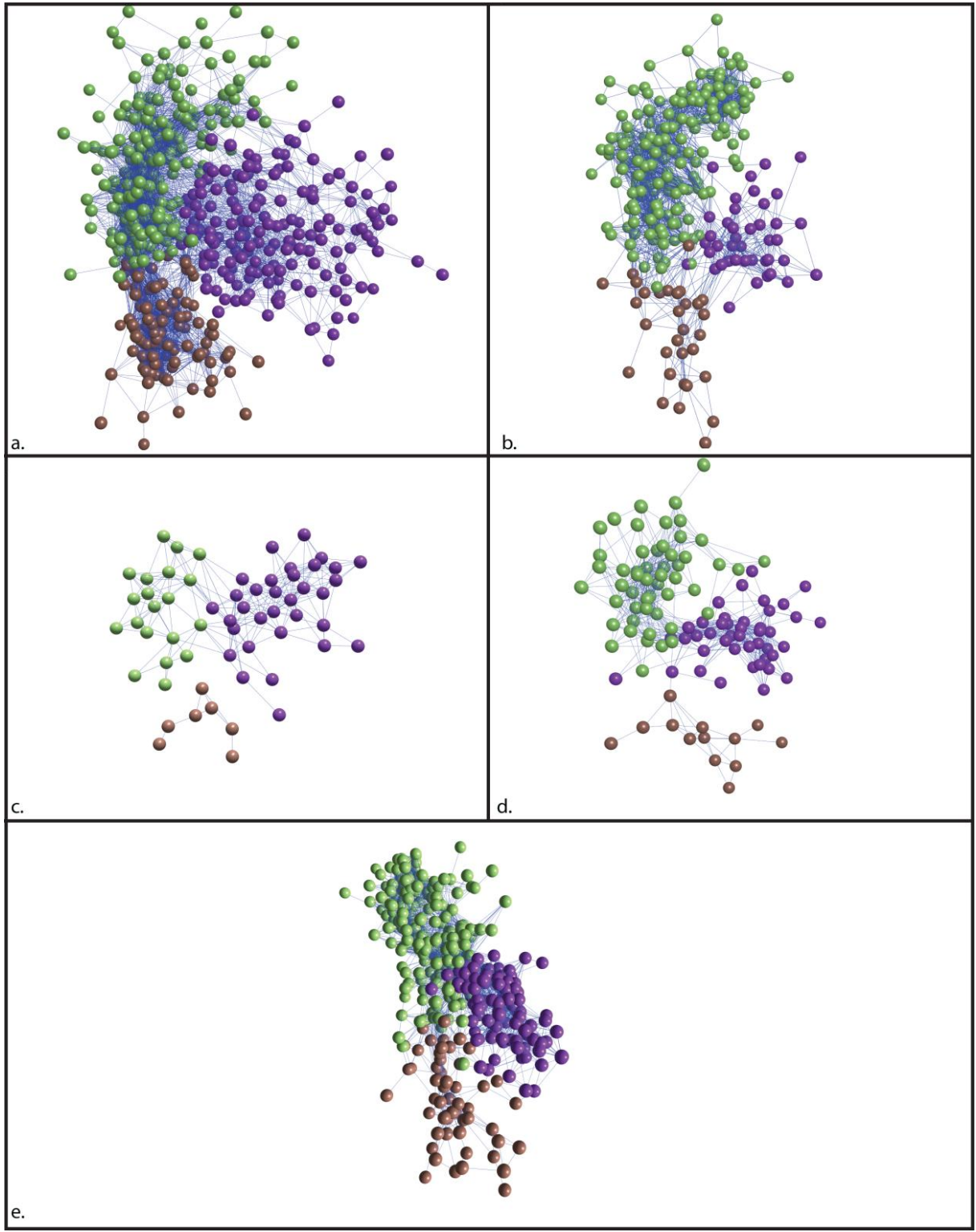

**S2. Supplementary symptom networks.**

The network graph created in Biolayout 3D Express, which incorporated data from all 1045 questionnaires and is displayed as *Figure 3*, was reproduced using differing questionnaire inclusion criteria. These all produced at least two distinct clusters when clustered using a MCL inflation value of 1.4. (A) includes questionnaires from Apex 2 Expedition subjects only (n=869); (B) includes only questionnaires from subjects not taking either sildenafil or antioxidant supplementation (n=523); (C) includes questionnaires from Kilimanjaro subjects only (n=176); (D) includes only questionnaires from subjects at a single time point (day 3 of the Apex 2 expedition, and all Kilimanjaro participants, n=269). (E) includes Apex 2 questionnaires, in which the sleep score from the following night was used in place of that from the preceding night (n=625).

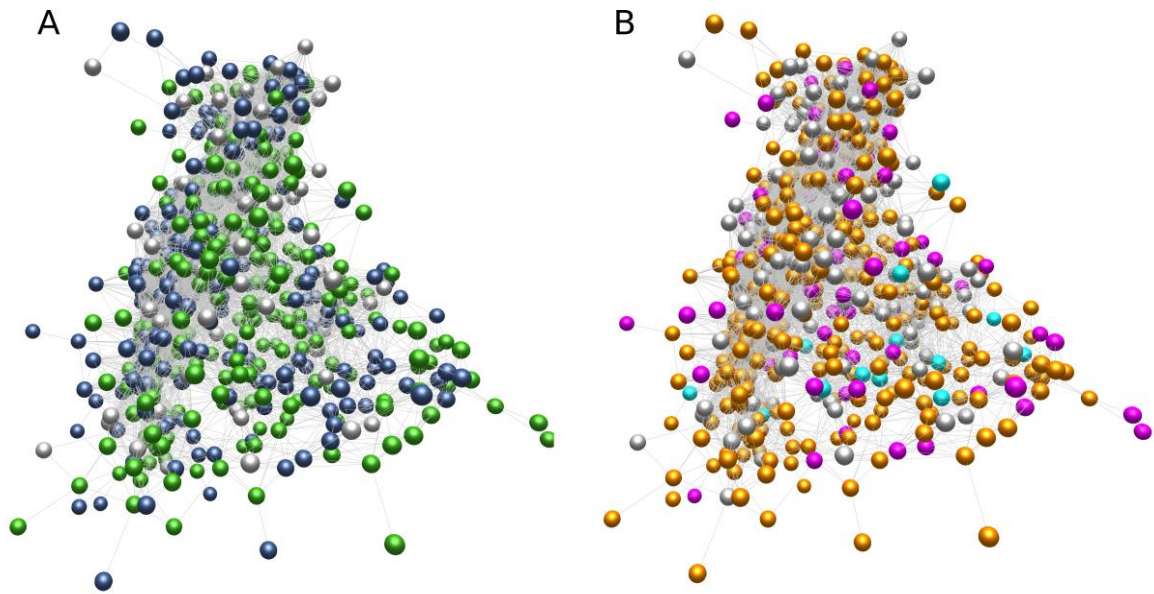

**S3. Age and sex distribution in symptom network.**

Each node (coloured sphere) represents a VAS questionnaire, connected by weighted lines, which represent correlations between similar symptom profiles. Nodes are connected with each other if the Pearson correlation coefficient between them exceeds 0.95. (A) Sex distribution of nodes, with questionnaires completed by males denoted by green nodes, those by females by blue, and missing demographic data by grey nodes; (B) Age distributions of nodes, with questionnaires completed by under 21 year olds represented by orange nodes, 22-25 year olds by pink nodes, and those completed by over 26 years old by cyan. Missing data are represented by grey nodes.